\documentclass[prc,showpacs,twocolumn]{revtex4}

\usepackage{graphicx}
\def\pt{$p_T$}
\def\dis{distribution}

\begin{document} 

\title 
{Azimuthal Anisotropy:  Ridges, Recombination\\ and Breaking of Quark Number Scaling}
\author
 {Charles B.\ Chiu$^1$, Rudolph C. Hwa$^2$ and C.\ B.\ Yang$^{2,3}$}
\affiliation
{$^1$Center for Particle Physics and Department of Physics, University of Texas at Austin, Austin, TX 78712, USA\\
$^2$Institute of Theoretical Science and Department of
Physics, University of Oregon, Eugene, OR 97403-5203, USA\\
$^3$Institute of Particle Physics, Hua-Zhong Normal
University, Wuhan 430079, P.\ R.\ China}

\begin{abstract} 
Azimuthal anisotropy is studied by taking into account the ridges created by semi-hard scattering, which is sensitive to the initial spatial configuration in non-central heavy-ion collisions. No rapid thermalization is required. Although hydrodynamics is not used in this study, the validity of hydrodynamical expansion is not excluded at later time after equilibration is achieved. Phenomenological properties of the bulk and ridge behaviors are used as inputs to determine the elliptic flow of pion and proton at low $p_T$. At intermediate $p_T$ the recombination of shower partons with thermal partons becomes more important. The $\phi$ dependence arises from the variation of the in-medium path length of the hard parton that generates the shower. The $p_T$ dependence of $v_2$ is therefore very different at intermediate $p_T$ compared to that at low $p_T$. Quark number scaling of $v_2$ is shown to be only approximately valid at low $p_T$, but is broken at intermediate $p_T$, even though recombination is the mechanism of hadronization in all $p_T$ regions considered.
 
 \end{abstract}
\maketitle
\section{Introduction}

In relativistic heavy-ion collisions the subjects of transverse momentum ($p_T$) distribution and azimuthal anisotropy have been given prominent attention both experimentally and theoretically from the very beginning \cite{ja}-\cite{ja3}.  In the second half of this decade details of the properties of elliptic flow have continued to be studied experimentally with greater accuracy \cite{ja4}-\cite{yb}, but there is a waning of further theoretical development, at least in the light quark sector.  The hydrodynamical model at low $p_T$ \cite{dt}-\cite{pk2} and the recombination-coalescence model at intermediate $p_T$ \cite{vg}-\cite{vg2} have described the data on elliptic flow so well that little room seems to exist for further improvement.  In particular, quark number scaling (QNS) is a property that has been scrutinized to the point where it has been regarded as a strong evidence for the partonic degree of freedom before hadronization and for recombination at low $p_T$.  In this paper we study the problem of azimuthal anisotropy in the framework of our version of the recombination model  \cite{rh} in which we consider thermal and shower partons.  We take into account the ridges at low $p_T$ generated by semi-hard scattering near the surface and the shower partons at higher $p_T$.  We calculate the second harmonic $v_2$ for pion and proton, as well as for a light quark, and show that, when $p_T$ is extended to the intermediate region, there is significant departure from the QNS result suggested by simple consideration of quark recombination \cite{dm,vg2}.

We state at the outset that we do not have a dynamical description of the time evolution of the expanding system.  We do not use hydrodynamics explicitly, but by considering thermal distribution of parton just before hadronization we imply the validity of hydrodynamical expansion at some point in the evolutionary history.  However, we do not subscribe to the applicability of hydrodynamics at early time, such as at $\tau_0 = 0.6$ fm/c \cite{pk,pk2}.  Rapid thermalization has not been shown to be the consequence of any dynamical process that is firmly grounded and commonly accepted.

It has been pointed out in Ref.\ \cite{rh2} that there exists an alternative mechanism to relate the spatial asymmetry at early time to the momentum anisotropy at late time, without relying on the assumption of fast equilibration.  That mechanism is semi-hard scattering, soft enough to have high probability of occurrence, but hard enough to take place at $\tau < 0.2$ fm/c.  The phenomenology associated with such processes may be termed ``ridgeology,'' which is the study of the properties of ridges that have been found to accompany jets, even when the initiating jets are weak and the peak-to-ridge ratio is small \cite{jp,jb}.  In this paper we develop further the study of elliptic flow based on ridges at low $p_T$ and shower partons at intermediate $p_T$.

The shower partons used in our model are defined at the hadronization scale so that the recombination of two types of shower partons in the same jet reproduces the fragmentation functions (FF) of a hard parton to a specific meson \cite{rh3}.  Similarly, three quarks in the shower can recombine to form a baryon in agreement with the baryon FF without adjusting any free parameters \cite{rh4}.  A shower parton can also recombine with a thermal parton in the immediate vicinity of the jet (but not belonging to a part of the jet), both being soft enough to undergo hadronization.  It has been shown in \cite{rh} that such thermal-shower (TS) recombination predominate in the region $3<p_T<6$ GeV/c.  Our concern for $v_2(p_T)$ in this paper will be for $0 < p_T<6$ GeV/c.

Above $p_T = 6$ GeV/c, SS (or SSS) recombination becomes important; that process is equivalent to fragmentation, since that is how the shower parton distributions are determined in the first place \cite{rh3}.  With that duality of fragmentation and recombination in mind, it is easy to see that QNS cannot be valid at high $p_T$, since jet quenching responsible for  $v_2(p_T)$ involves only one hard parton, not two (or three) semi-hard partons.  Although we do not consider the region $p_T >6$ GeV/c, there is still the question:  at what point does QNS begin to break down.  Our study here shows that the breaking of QNS begins at the transition point between TT and TS recombination for pion and between TTT and TTS+TSS for proton.  The reason is in the nature of TS and TTS recombination, a bad approximation of which can lead to a simplistic formula that falsely suggests QNS.
Data that seem to support QNS are all for minimum bias, and do not reach the upper region of intermediate $p_T$.

In Sec.\ II we give the general formulation of how azimuthal anisotropy can arise from ridges without any details on the $p_T$ dependence.  Elliptic flow is then calculated for pion and proton production at low $p_T$ in Sec.\ III, where only thermal partons are considered.  In Sec.\ IV the study is extended to intermediate $p_T$, where the contribution from shower partons is included.  The breaking of quark number scaling is investigated in Sec. V.  The final section contains the conclusion of this work.

\section{Azimuthal anisotropy arising from ridges}

Let us first review our approach to elliptic flow at low $p_T$ without rapid thermalization \cite{rh2}.  Instead of assuming the meaningfulness of thermodynamical quantities, like pressure and temperature, at early time, we recognize that hard scattering of partons can occur at all virtuality $Q^2$, with increasing probability at lower and lower $Q^2$, and that when the parton transverse-momentum, $k_T$, is around 2 - 3 GeV/c, the rate of such semi-hard scattering can be high, while the time scale involved is low enough ($\sim 0.1$ fm/c) to be sensitive to the initial spatial configuration of the collision system.  When such scattering occurs near the surface of the overlap region in the transverse plane, each semi-hard parton creates a ridge in $\Delta \eta$ and $\Delta \phi$ \cite{ja6}.  When triggers are used, ridges have been found in the associated-particle distribution on the near side with trigger momentum $p^{\rm trig}_T > 4$ GeV/c \cite{ja6}.  More recently, $p^{\rm trig}_T$ has been reduced to as low as 2.2 GeV/c and $p^{\rm assoc}_T$ as low as 1.5 GeV/c, where the ridge yield significantly dominates over the yield of the peak that sits above the ridge \cite{jb}.  It suggests that ridges due to the scattering of low-$x$ partons $(<0.03)$ are abundantly produced even if triggers are not used to select events to examine their properties.  The effect of such ridges on both the $p_T$ and $\phi$ dependences of the produced particles should not be ignored.

If the semi-hard scattering occurs in the interior of the dense medium, the  energy of the scattered partons is dissipated in the medium and contributes to the thermalization of the bulk. That process may take some time to complete, a likelihood that is acceptable in the approach adopted here, since we have no need to require thermalization to be fast.  If the semi-hard scattering occurs near the surface of the medium, one of the scattered partons may be directed outward and lose energy to the medium on its way out.  The enhanced thermal partons near the jet trajectory can recombine and form hadrons in the ridge.  This interpretation of the ridge has been applied successfully to triggered events \cite{cc}, and provides a resolution to the $\Omega$ puzzle \cite{rh5,cc2}.  The same mechanism of ridge formation is, however, also valid, if no trigger is used.  Although the variables $\Delta \eta$ and $\Delta \phi$, usually defined relative to the trigger momentum, would be meaningless without a trigger, the existence of ridge in the difference variables $\eta_\Delta $ and $\phi_\Delta$ in autocorrelation that used no triggers has been well established \cite{ja6,ja7}.

The direction of a scattered parton is random, but after it is averaged over all events, the average direction of all outward partons near the surface is normal to the surface.  Since low-x partons are copious and the rate of semi-hard scattering with $k_T <3$ GeV/c is not significantly suppressed, the semi-hard emitters can form a layer along the surface of the overlap region prescribed by the geometry at the initial time of collision.  The formation of the ridge of hadrons takes some time for the transverse expansion to complete, but the directions in which the ridge partons flow are determined by spatial configuration at early time.  We do not rule out the applicability of hydrodynamics at some point of the expansion process when equilibration is established.  However, fast thermalization is not needed if semi-hard scattering can initiate the anisotropic expansion.

Details about ridge formation are rather complicated, especially if a model is to be successful in reproducing the very recent data on the ridge yield as a function of the trigger azimuthal angle relative to the reaction plane  \cite{af}. A full treatment of that problem is a separate subject of its own \cite{ch3}, and is unsuitable for inclusion here; furthermore, we do not use triggers in the study of inclusive distributions from which $v_2$ is to be determined. However, aspects of that problem are needed to demonstrate the process of averaging over the jet direction of the semi-hard parton. We describe in the Appendix our model calculation of the ridge \dis\ in the azimuthal angle $\phi$, and show how the result can be represented by a simple box approximation. We take that approximation as the starting point here and proceed to the calculation of $v_2(p_T)$.

The overlap region in the transverse plane for two nuclei of radius $R_A$ at impact-parameter $b$ apart is, assuming simple geometry with sharp boundaries, the almond-shaped area bounded by two circular arcs whose maximum angle is $\Phi$, where
 \begin{eqnarray}
\cos\Phi=\hat{b} \equiv b/2R_A  \ ,   
\label{1}
\end{eqnarray}
and the angle $\phi$ within the arcs satisfies $\phi\in{\cal R}$, which is a set of angles defined by 
\begin{eqnarray}
|\phi| < \Phi \, \qquad{\rm and}\, \qquad |\pi - \phi| < \Phi  .   
\label{2}
\end{eqnarray}
It should be emphasized that the almond-shaped region is relevant for the consideration of the initial problem of semi-hard scattering whose time scale is short, although hadrons in the ridge are formed later when the elliptic geometry is more pertinent.
In the Appendix we show that the ridge ($R$) \dis\ in $\phi$ can be represented by the box approximation
\begin{eqnarray}
R(p_T, \phi) =  R(p_T) \Theta (\phi)    \ ,   
\label{3}
\end{eqnarray}
where
\begin{eqnarray}
\Theta (\phi) = \theta (\Phi - |\phi|) + \theta (\Phi - |\pi - \phi|)    \ .   
\label{4}
\end{eqnarray}
The bulk $(B)$ medium has no $\phi$ dependence and will be denoted by $B(p_T)$.  The single-particle distribution at low $p_T$ is then
\begin{eqnarray}
{dN \over p_T dp_T d\phi}=  B(p_T) +   R(p_T) \Theta (\phi)  \ .  
\label{5}
\end{eqnarray}
This is a simple expression in closed form that enables us to calculate $v_2(p_T)$ analytically.

The second harmonic in the $\phi$ distribution is 
\begin{eqnarray}
v_2 (p_T) = \left<\cos 2 \phi\right > = {\int^{2 \pi}_0 d\phi\ \cos 2 \phi\, dN/ p_T dp_T d\phi \over \int^{2 \pi}_0 d\phi\ dN/ p_T dp_T d\phi}  \ .   
\label{6}
\end{eqnarray}
When Eq. (\ref{5}) is used in the above, we obtain
\begin{eqnarray}
v_2 (p_T, b ) = {\sin 2 \Phi (b)  \over \pi B(p_T)/R(p_T) + 2 \Phi (b)}  \ .   
\label{7}
\end{eqnarray}
  If the first term in the denominator is much larger than the second, as we shall show below for low $p_T$, then we have the even simpler formula
\begin{eqnarray}
v_2 (p_T, b ) \simeq  { R(p_T)  \over \pi B(p_T)}  \sin 2 \Phi (b)  \ ,   
\label{8}
\end{eqnarray}
where the $p_T$ and $b$ dependences are factorized.  Thus the centrality dependence at fixed $p_T$ is essentially specified by $\sin 2 \Phi (b)$, and is in accord with the data at low $p_T$ \cite{rh2}.  We shall show more detailed properties of the centrality dependence below.
Equation (\ref{8}) is the analytic result that represents in a simple approximation the consequence of considering the effects of semi-hard scattering instead of fast thermalization.

\section{Elliptic flow at low $p_T$}

Let us now consider the low-$p_T$ behaviors of $B(p_T)$ and $R(p_T)$.  The parton distribution in transverse momentum $q_T$  of the bulk medium  has the thermal behavior \cite{rh}
\begin{eqnarray}
q_0 {dN^B_q\over dq_T d\phi} = C q_Te^{-q_T/T}
\label{9}
\end{eqnarray}
at low $q_T$. 
This distribution includes the effect of semi-hard partons created in the interior of the medium; they lack the energy to reach the surface to get out. The energy lost to the medium is thermalized and contributes to the value of the effective temperature $T$. The time it takes for the weak jets to thermalize need not be short; it can take a significant part of the expansion phase of the whole system, if necessary. If the semi-hard scattering occurs near the surface, one of the scattered partons may emerge, while the recoil parton directed inward gets thermalized. The emerging semi-hard partons along the surface lose extra energy to the medium in addition to those others that cannot escape. Thus there is an enhancement over the bulk, for which the thermal distribution has the same form but with a higher inverse slope $T'$
\begin{eqnarray}
q_0 {dN^{B+R}_q\over dq_T d\phi} = C q_Te^{-q_T/T'} , \qquad\phi \ {\in} \  \cal R
\label{10}
\end{eqnarray}
The difference between Eqs.\ (\ref{9}) and (\ref{10}) is the ridge effect. 
Note that the normalization factor $C$ is the same in Eqs.\ (\ref{9}) and (\ref{10}) because there is no difference between the weak jets that fail to emerge from the surface and those that barely emerge with negligible energy to develop any additional enhancement at $q_T=0$. It is only when the semi-hard scattering occurs sufficiently near the surface that a substantial ridge can be formed with non-vanishing $q_T$. The  two equations above are not derived from any dynamical equation of time evolution, but are to be inferred from the data on particle distribution in \pt\ after the quarks hadronize by recombination.  Clearly, the use of thermal distributions implies equilibration before hadronization, but it need not be accomplished at early time.  It is not necessary for us to specified how long the equilibration time is, since Eqs.\ (\ref{9}) and (\ref{10}) may be regarded as phenomenological input with $T$ and $T'$ to be determined from data.

To derive the pion and proton distributions from Eqs.\ (\ref{9}) and (\ref{10}) the formalism in \cite{rh} is to be used.  However, in \cite{rh} only central collision is considered, for which the quark distributions are assumed factorizable before recombination.   Here for the study of $\phi$ anisotropy all centralities must be considered, and the assumption of factorizability of $uud$ distribution for the production of proton in peripheral collision (where thermal parton density is low) is questionable.  We shall treat such complications in a later section.  For now, let us ignore the issue of non-factorizability and proceed as in \cite{rh} for central and mid-central collisions so as to show the connection between the ridge effect and elliptic flow.

\subsection{Pion}

Starting from Eq.\ (\ref{9}), using valon distribution in the recombination function, and neglecting pion mass, we obtain in \cite{rh} the pion distribution due to $TT$ recombination
\begin{eqnarray}
B_{\pi}(p_T) = {dN^{B}_{\pi} \over p_Tdq_T d\phi} = {C^2 \over 6} e^{-p_T/T} 
\label{11}
\end{eqnarray}
for the bulk medium at any $\phi$.  The normalization factor $C^2$ will be canceled at low $p_T$, so it need not be specified here.  $T$ will be discussed below.  It should be recognized that the form in Eq.\ (\ref{9}) is chosen so as to yield the exponential distribution in (\ref{11}).  Starting from  Eq.\ (\ref{10}) we obtain similarly for $\phi \in {\cal R}$
\begin{eqnarray}
B_{\pi}(p_T) + R_{\pi}(p_T,\phi) = {dN^{B+R}_{\pi} \over p_T dq_T  d\phi} = {C^2 \over 6} e^{-p_T/T'}  .
\label{12}
\end{eqnarray}
  While Eqs.\ (\ref{11}) and (\ref{12}) are consequences of (\ref{9}) and (\ref{10}) for any $p_T$, their phenomenological application is useful only for low $p_T$ where recombination with shower partons is relatively unimportant.  However, for $p_T \approx 0$  the hadronic mass cannot be ignored, even for pion, let alone proton.  The recombination model we use is based on a formalism for large momenta \cite{kd}- \cite{rh7}, where momentum fractions are meaningful.  Thus it is an assumption that the model remains quantitatively valid at low $p_T$, an extrapolation that can be made more acceptable if the mass effect can be taken into account.  To that end, we adopt the ansatz that $p_T$ in Eqs.\ (\ref{11}) and (\ref{12}) are to be replaced by the transverse kinetic energy $E_T$
  \begin{eqnarray}
E_T (p_T) = m_T - m_0, \qquad m_T = \left(p_T^2 + m_0^2 \right)^{1/2} \ ,
\label{13}
\end{eqnarray}
where $m_0$ is the hadron rest mass.  We then have
\begin{eqnarray}
B_{\pi}(p_T) =  {C^2 \over 6} e^{-E_T(p_T)/T} 
\label{14}
\end{eqnarray}
\begin{eqnarray}
B_{\pi}(p_T) + R_{\pi}(p_T,\phi) = {C^2 \over 6} e^{-E_T(p_T)/T'},\quad  \phi   \in  {\cal R},
\label{15}
\end{eqnarray}
where $T$ and $T'$ remain to be determined from data.  Subtracting Eq.\ (\ref{14}) from (\ref{15}) results in (\ref{3}), where $R_{\pi}(p_T)$ has now the concrete form 
\begin{eqnarray}
R_{\pi}(p_T)  = {C^2 \over 6} e^{-E_T(p_T)/T'} \left( 1- e^{-E_T(p_T)/T''}\right) \ ,
\label{16}
\end{eqnarray}
\begin{eqnarray}
{1 \over T''}  = {1 \over T}- {1 \over T'} = {\Delta T \over TT'}, \qquad \Delta T = T' - T \ .
\label{17}
\end{eqnarray}
For the values of $T$ and $T'$, we examine the data from STAR that has focused on ridgeology \cite{jp}.  It has been found that the ridge distribution in $p_T^{\rm assoc}$ is nearly exponential and remains essentially the same for a wide range of trigger momentum $p_T^{\rm trig} > 4$ GeV/c.  The minimum value of $p_T^{\rm assoc}$ measured is, however, 2.2 GeV/c, too high to exhibit the small $E_T$ behavior in Eq.\ (\ref{16}).  Furthermore, it is also shown in Ref.\ \cite{jp} that the $p/ \pi$ ratio in the ridge for $4 < p_T^{\rm trig} <5$ GeV/c is about 4.5.  It means then that the data on ridge in \cite{jp} for unidentified charged particles cannot be used to determine $T'$ in Eq.\ (\ref{16}) for pion.  It should also be pointed out that in \cite{jp} the ridge slope is compared to the inclusive slope of charged particle distribution for the same range of $p_T$ as for $2.2 < p_T^{\rm assoc} < 4$ GeV/c.  That is the region in which the $p/ \pi$ ratio of the inclusive single-particle distributions is approximately 1.    So no information about the pion $T$ for the bulk can be obtained from Ref.\  \cite{jp} either.

For central collisions we have $b \simeq 0$ and $\Phi \simeq \pi/2$, so $B(p_T) + R(p_T,\phi) $ is for all $\phi$ without anisotropy.  The single-particle distribution is therefore a measure of $B + R$, not $B$ alone.  Semi-hard scattering is present in all events so ridges contribute to $dN/p_T dp_T$ whether or not triggers are used to select jet events.  Thus Eq.\ (\ref{15}) can be applied to the inclusive data for identified pions to determine $T'$.  The data are not exactly exponential for all low $p_T$.  For $p_T < 1$ GeV/c there are pions from resonance decays.  We extract the value of $T'$ from the region  $p_T > 1$ GeV/c, in which the ridge particles from the enhanced thermal source make definitive contribution.  Using the data in Ref. \cite{sa} we fit the $\pi^+$ distribution in $E_T$ at 0-5\% centrality for $1 < E_T < 3$ GeV and get $T' = 0.3 $ GeV.  If TT recombination were dominant in peripheral collisions, we need only go to the $\pi^+$ distribution at 80-92\% centrality and determine $T$, since the ridge contribution is small when $\Phi$ is small.  But that is not the case.  At large $b$ the thermal source is weak, and SS recombination dominates.  It is shown in Ref.\ \cite{rh8} that SS $>$ TT for $p_T > 2$ GeV/c at 80-92\%, as in $pp$ collisions.  Without direct experimental guide to determine $T$ for $B_{\pi}(p_T)$, we proceed by adopting $\Delta T = 45$ MeV in common with what we shall be able to obtain in the study of the proton case to be described below; it is consistent with the range of values suggested experimentally in \cite{jp} and used in \cite{rh2}.  Using Eq.\ (\ref{17}) we then obtain $T = 0.255$ GeV and 
\begin{eqnarray}
 T''_{\pi} = 1.7\ {\rm GeV}
\label{18}
\end{eqnarray}
for the pion.

The ratio $B_{\pi}(p_T)/R_{\pi}(p_T)$ has a simple form when Eq.\ (\ref{16}) is rewritten as 
\begin{eqnarray}
R_{\pi}(p_T)  = {C^2 \over 6} e^{-E_T(p_T)/T} \left(e^{E_T(p_T)/T''}-1\right) \ ,
\label{19}
\end{eqnarray}
so that the prefactors cancel and we have
\begin{eqnarray}
{B_{\pi}(p_T) \over R_{\pi}(p_T)}  = {1 \over e^{E_T(p_T)/T''}-1 }  \ ,
\label{20}
\end{eqnarray}
which depends only on $T''$.  We assume the validity of this equation for all $p_T < 2$ GeV/c where TT recombination is valid.  When $p_T$ is small, Eq.\ (\ref{20}) can be approximated by $T''/E_T$, so 
Eq.\ (\ref{8}) has the simple expression
\begin{eqnarray}
v^{\pi}_2 (p_T, b) = {E_T (p_T) \over \pi T''} \sin 2\Phi(b) \ .
\label{21}
\end{eqnarray}

\begin{figure}[tbph]
\includegraphics[width=0.45\textwidth]{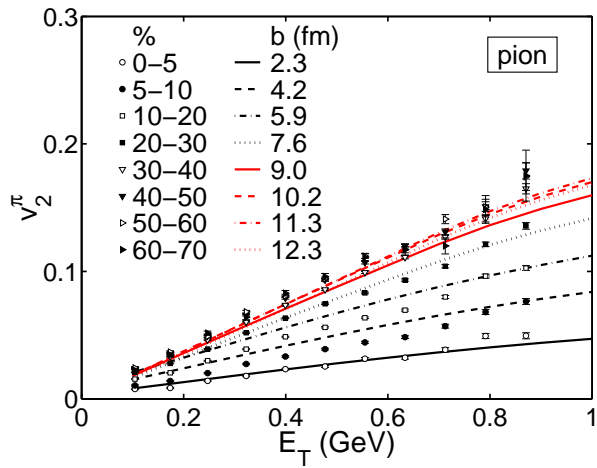}
\caption{(Color online) Comparison of calculated $v_2^{\pi}$ with data for Au+Au collisions at 200GeV \cite{ja4} for 8 centrality bins whose corresponding values of $b$ are shown in the legend.}
\end{figure}
\noindent The initial slope is
\begin{eqnarray}
\left.{\partial v^{\pi}_2\over \partial E_T}\right|_{E_T = 0} = {1 \over \pi T''_{\pi}} \sin 2 \Phi (b) \ ,
\label{22}
\end{eqnarray}
which is dependent on only $T''_{\pi}$ apart from the geometrical factor $\sin 2 \Phi (b)$.  That property is in good agreement with the data \cite{ja4}, as shown in Fig.\ 1.  Equation (\ref{22}) is a distinctive feature of $v_2$ that is driven by ridges.  For non-vanishing values of $E_T$ we use the full expression of Eq.\ (\ref{20}) in (\ref{7}), obtaining the result shown in Fig.\ 1 for $p_T < 1$ GeV/c.  Deviation from linearity is perceptible.  There is some discrepancy between our results and the data, the most noticeable region being around $E_T \sim 0.65$ GeV, but taken as a whole the overall agreement with the data is good.  That is remarkable, since we have not adjusted any free parameters in order to fit.

It is worthwhile to digress and explain why our model that depends on the initial geometry of the medium can produce satisfactory $v_2(p_T)$, whereas earlier model based on surface emission without hydrodynamical expansion failed \cite{hkh}.  In our case, although the average direction of the semi-hard partons are emitted normal to the surface of the initial configuration, the hadrons in the ridges are produced after expansion. The late emission of hadrons introduces a scale in $R_\pi(p_T)/B_\pi(p_T)$ in the form of $T''$, which controls the normalization in Eq.\ (\ref{22}). The early emission of semi-hard partons determines the $\phi$ dependence through $\Phi(b)$ in agreement with data. In the case of \cite{hkh} the only scale is the temperature $T=140$ MeV without expansion, so $v_2(p_T)$ saturates at around $p_T\approx 0.2$ GeV/c in disagreement with data. Furthermore, it is not detailed enough to give centrality dependence. The importance of hydrodynamical expansion is pointed out in \cite{hkh}, on which we have no disagreement. In summary, ridge physics is not a simple surface-emission problem because it comprises both early-time parton emission and late-time hadron formation, and therefore has more dynamical content to result in the proper azimuthal anisotropy.

\subsection{Proton}

For proton production in central Au-Au collision we have obtained  for TTT recombinations \cite{rh}
\begin{eqnarray}
{dN_p \over p_T dp_T} = A {p_T^2 \over p_0}e ^{-p_T/T}
\label{23}
\end{eqnarray}
where
\begin{eqnarray}
A = {C^3 \over 6} {B(\alpha + 2, \gamma + 2) B(\alpha + 2, \alpha+\gamma + 4) \over B(\alpha + 1, \gamma + 1) B(\alpha + 1, \alpha+\gamma + 2) } \ ,
\label{24}
\end{eqnarray}
$\alpha = 1.75$ and $\gamma = 1.05$.  These beta functions come from the recombination function of proton, which depend on valon distribution characterized by $\alpha$ and $\gamma$ \cite{rh9}.  The factor $p^2_T$ arises from the integration over the momenta of the quarks that recombine.  The factor $p_0$ comes from the invariant distribution $p_0dN/d^3p$ on the left-hand side.  When applied to $p_T > 2$ GeV/c at mid-rapidity, the $p^2_T/p_0$ factor was approximated by $p_T$ in \cite{rh}.  Here we want to extend Eq.\ (\ref{23}) to lower $p_T$, still at $y \approx 0$, so to take the mass effect into account we rewrite the equation in the form 
\begin{eqnarray}
B_p (p_T) = A {p^2_T \over m_T} e^{-E_T (p_T)/T}.
\label{25}
\end{eqnarray}
Similarly, for bulk + ridge we have 
\begin{eqnarray}
B_p (p_T) +  R_p (p_T,\phi)  = A {p^2_T \over m_T} e^{-E_T (p_T)/T'}, \   \phi \in {\cal R}.
\label{26}
\end{eqnarray}
The ridge solution is then for $\phi \in {\cal R}$
\begin{eqnarray}
R_p (p_T)  = A {p^2_T \over m_T} e^{-E_T (p_T)/T'}\left(1 - e^{-E_T (p_T)/T''} \right) .
\label{27}
\end{eqnarray}

When the exponential factor is in terms of $p_T$, as shown in Eq.\ (\ref{23}), the value of $T$ should be the same as that in (\ref{11}) for pions, as well as that in (\ref{9}) for the quarks.  That is because of the $\delta \left(p_T - \sum_i q_{i_T} \right)$ function in the recombination function that preserves the exponential behavior  from quarks to pions to proton.  Experimentally, it has been found that the $p_T$ distributions of $\pi^+$, $K^+$ and $p$ have the same inverse slope in their exponential behavior for $1.5 < p_T <3$ GeV/c (see Fig.\ 5 of \cite{sa}), thus confirming the result of the recombination model.  However, the same data, when plotted in terms of $E_T$, show different slopes (see Fig.\ 9 of \cite{sa}).  As stated in the Sec. III.A above, the slopes determined from the $E_T$ distribution for 0-5\% centrality is the value of $T'$.  We obtain from the proton distribution in \cite{sa} $T' = 0.35$ GeV.  It is now possible to extract some information from the ridge distribution in Ref.\ \cite{jp} which is for 0-10\% centrality.  The lowest $p_T^{\rm trig}$ range is $4<p_T^{\rm trig} < 5$ GeV/c for which the $p/\pi$ ratio is 4.5.  Assuming dominance by proton, we use Eq.\ (\ref{27}) to fit the data by varying $\Delta T$.  
Since the normalization of the ridge distribution in \cite{jp} is dependent on $p_T^{\rm trig}$, whereas Eq.\ (\ref{27}) makes no explicit reference to $p_T^{\rm trig}$, we focus only on the $p_T$ dependence of the particles in the ridge and adjust the normalization to fit.
The best fit is for $\Delta T = 45$ MeV, for which the result is shown in Fig.\ 2.  The data points are outside the $p_T$ range where the dip occurs at low $p_T$; nevertheless, a good fit is achieved in the region where data exist.  It should be recognized that the factor $p^2_T$ in front of $\exp (-E_T/T')$ in  Eq.\ (\ref{27}), as well as the factor, $1 - \exp (-E_T/T'')$, after it together make the effective inverse slope to be greater than $T'$, where the data points are.  A measurement of the dip in the proton \dis\ in the ridge at small $p_T$ in Fig.\ 2 will serve to test the validity of our model.  With $\Delta T = 0.045$ GeV we obtain from Eq.\ (\ref{17}) $T = 0.305$ GeV, and 
\begin{eqnarray}
T''_p = 2.37\ {\rm GeV} 
\label{28}
\end{eqnarray}
for 0-10\% centrality.

\begin{figure}[tbph]
\includegraphics[width=0.45\textwidth]{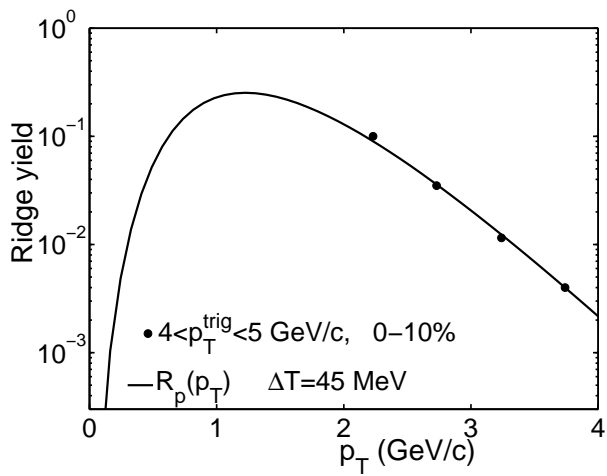}
\caption{Fit of the \pt\ \dis\ of charged particles in the ridge  \cite{jp} by Eq.\ (\ref{27}).}
\end{figure}

For centrality dependence we go to Ref.\ \cite{ka2} and find that, whereas the slope for pion is essentially independent of $c$ (defined as centrality in \%), that for proton (and antiproton) decreases with $c$.  Since the $E_T$ distribution of proton has a break in slope from $E_T < 1$ GeV to $> 1$ GeV, we choose to consider the tabulated slope for $\bar{p}$, which is independent of the $E_T$ regions and in our view should be the same as for $p$.  We find that (identifying $T'_p = T'_{\bar{p}}$)
\begin{eqnarray}
T_p' = 0.35\,(1 - 0.5 \hat{c})\ {\rm GeV}, \qquad \hat{c} = c/100
\label{29}
\end{eqnarray}
gives a good fit of the data as shown in Fig.\ 3.
The reason for the centrality dependence is that for baryon production the three quarks (or antiquarks) needed for recombination become less independent of one another at larger $c$, where less thermal partons are generated in the nuclear collision.  We know that in very peripheral collision the system should approach that of $pp$ collision and few thermal partons are created for baryon production.  When the $uud$ joint distribution is not factorizable, our model calculation of $p$ production that is based on factorizable quark distribution cannot be applied directly.  Equation (\ref{29}) is the simplest way to summarize the $c$ dependence of the inclusive distribution.

\begin{figure}[tbph]
\includegraphics[width=0.45\textwidth]{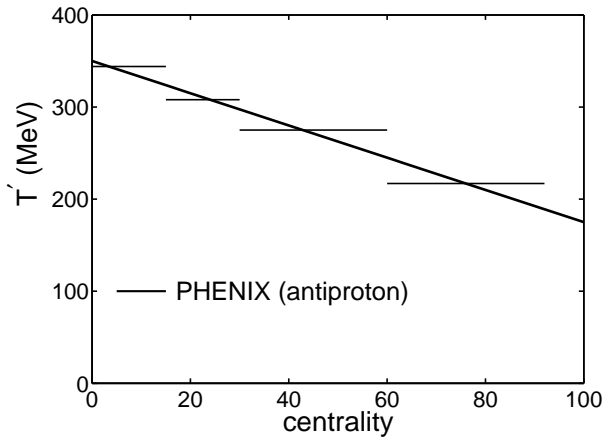}
\caption{Fit of inverse slopes of $\bar p$ production for five centrality bins \cite{ka2}.}
\end{figure}

Fixing $\Delta T$ at 45 MeV, we have
\begin{eqnarray}
T_p = 0.305 - 0.175 \hat{c}\ {\rm GeV} \ .
\label{30}
\end{eqnarray}
$T''_p$ follows simply from Eq.\ (\ref{17}).  The result can be well approximated by a convenient formula
\begin{eqnarray}
T''_p= 2.37\,(1 - 1.05 \hat{c} + 0.26 \hat{c}^2)\ {\rm GeV} \ . 
\label{31}
\end{eqnarray}
The consequence on $v_2$ can now be calculated as in the case of pion, using Eq.\ (\ref{21}) for low $E_T$.  The initial slope for proton is then
\begin{eqnarray}
\left.{\partial v^p_2\over \partial E_T}\right|_{E_T = 0} = {1 \over \pi T''_p (\hat c)} \sin 2 \Phi (b)\ ,
\label{32}
\end{eqnarray}
Since $T''_p \neq T''_{\pi}$, the initial slopes for proton and pion are not the same.  That seems to violate quark number scaling, since dividing $v_2$ and $E_T$ by the same $n_q$ does not change the initial slopes.  However, the empirical evidence for QNS is only for minimum bias data \cite{ba,at,aa}.  If we evaluate $T''_p (\hat{c})$ at $c = 30$\%, we find  $T''_p (0.3)$ to be very nearly 1.7 GeV in agreement with Eq.\ (\ref{18}).  Thus our result is in accord with the experimental evidence for QNS at low $E_T$, but also indicates violation of QNS in central and peripheral collisions.  At higher $E_T$ there are other reasons for more serious breaking of QNS; that will be discussed in a later section.

\begin{figure}[tbph]
\includegraphics[width=0.45\textwidth]{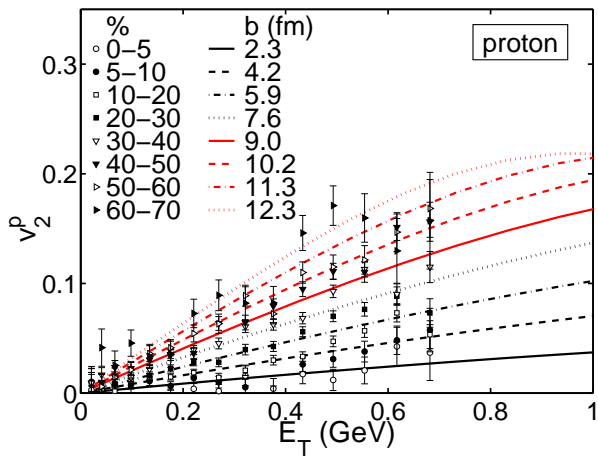}
\caption{(Color online) Comparison of calculated $v_2^{p}$ with data for Au+Au collisions at 200GeV \cite{ja4} for 8 centrality bins whose corresponding values of $b$ are shown in the legend.}
\end{figure}

At higher $E_T$, but still $<1$ GeV, $B_{\pi}(p_T) / R_{\pi}(p_T)$ has the same form as Eq.\ (\ref{20}), so $v^p_2(p_T,b)$  can be calculated using Eq.\ (\ref{7}).  The result is shown in Fig.\ 4 in rough agreement with the data that have large errors \cite{ja4}.  Note that whereas $v^{\pi}_2(p_T,b)$ saturates at around 40-60\% centrality, $v^p_2(p_T,b)$ continues to rise at more peripheral collisions.  That is because, as $b$ increases, $T''_p (c)$ monotonically decreases so that $\sin 2 \Phi (b)$ is not the only factor that determines the dependence on centrality, as is the case with pions.  The maximum of    
$\sin 2 \Phi (b)$ occurs at $\Phi = \pi/4$, corresponding to $\hat{b} = 1/\sqrt{2}$, or $\hat{c} = 1/2$.  Equations (\ref{7}), (\ref{20}) and (\ref{32}) indicate that the decrease of $T''_p (\hat c)$ causes $v^p_2 (E_T, b)$ to continue to increase beyond $\hat{b} = 0.7$ until the severe decrease of $\sin 2 \Phi (b)$ at large $\hat{b}$ brings it down.  We show in Fig.\ 5 the  $\hat{b}$ dependence (for $2R_A = 14.7$ fm) of 
$v^{\pi}_2 (p_T, \hat{b})$ and $v^p_2 (p_T, \hat{b})$ at fixed $p_T = 0.52$ GeV/c.  The characteristics of the data \cite{ja4} are well captured by the simple formula, Eq.\ (\ref{7}), for both pion and proton.

\begin{figure}[tbph]
\includegraphics[width=0.45\textwidth]{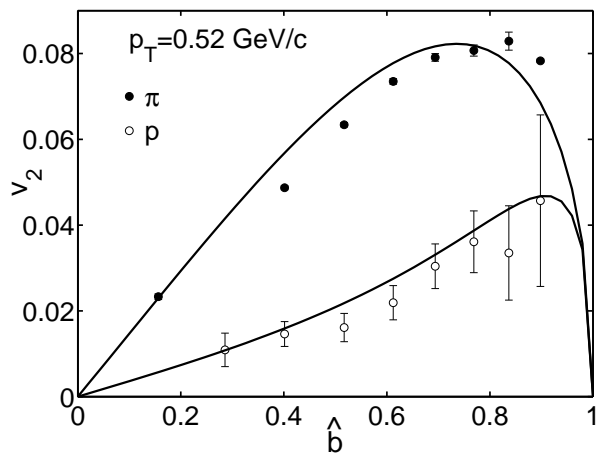}
\caption{Dependence of $v_2(p_T,b)$ on $\hat b=b/2R_A$ at $p_T=0.52$ GeV/c for pion and proton. Data are from Ref.\ \cite{ja4}. }
\end{figure}

\section{Azimuthal anisotropy at intermediate $p_T$}

As $p_T$ is increased to above 2 GeV/c, it is necessary to consider the role played by the shower partons \cite{rh}.  Both the ridge and shower partons are effects of jets, the former due to semi-hard partons, and the latter due to harder partons.  The transition from one to another is, of course, a continuous one.  Our recombination model was formulated in \cite{rh} at a time before the ridges were discovered.  In that model the three types of recombination are TT, TS and SS, where no ridges are considered in the thermal partons.  We now realize, as discussed in the preceding section, that the thermal distribution is $B+R$, since semi-hard scattering is always present.  The only effect of this realization is just to relabel $T$ in previous work by $T'$ now, as its value is determined from data.  As we proceed to consider TS recombination in this section, it is $T'$ that we shall use for the thermal partons.  The condition of $\phi \in {\cal R}$ in Eq.\ (\ref{10}) for $T'$ to be used is for TT recombination.  Now for TS recombination, there is a hard parton to generate the shower parton.  That hard parton may have any $\phi$.  Even if $\phi \not \in {\cal R}$, the energy loss of the hard parton can enhance the thermal partons near its trajectory, so the new $T'$ can depend on $b$ and $\phi$, characterizing the new ridge associated with hard parton. The $\phi$ dependence of the thermal partons is different from that of the shower partons. The latter is
 due to jet quenching of the hard parton that depends on path length in the medium.  This change of the origin of the $\phi$ anisotropy, when TT dominance is replaced by TS dominance at higher $p_T$, is the basic cause for the change of the nature of $p_T$ dependence of $v_2(p_T)$.
 
 It should be noted that in other recombination models the azimuthal anisotropy at intermediate and higher \pt\ has been studied already \cite{vg,rf}. Our approach here is different in the use of shower partons and in the inclusion of medium effect by treating TS recombination, whereas earlier studies considered fragmentation of hard partons as an additive component. The effect of jet quenching alone on $v_2$ for $p_T>2$ GeV/c in a simple geometrical study of the $\phi$ dependence is known to be too low \cite{es}.

\subsection{Hard Parton's $\phi$ Dependence}

The shower parton distribution $S^j_i (z)$ is the invariant probability of finding a parton of type $j$ with momentum fraction $z$ in a shower initiated by a hard parton of type $i$.  Its detail properties are described in Refs. \cite{rh3,rh4}, and its application to TS recombination in central collisions averaged over all $\phi$ is discussed in Ref.\ \cite{rh}.  We now consider $\phi$ dependence due to energy loss of the hard parton with varying path length in the dense medium.  Let us denote the distribution of hard parton $i$ emerging from the surface of the medium with transverse momentum $k$ at angle $\phi$ by
\begin{eqnarray}
\left.{dN^{\rm hard}_i \over kdkdyd\phi}\right|_{y = 0} =  F_i (k, \phi)\ .
\label{33}
\end{eqnarray}
Jet quenching degrades the hard-scattering momentum from the value $k'$ at the point of scattering to the emerging momentum $k$ by an amount $\Delta k$ that depends on the path length $\ell (\phi)$ in the medium.  The maximum length, $\ell _{\rm max}$ of a straight line at angle $\phi$ that passes through the origin of the almond-shaped overlap region in the transverse plane satisfies the equation \begin{eqnarray}
\left(\ell _{\rm max} \cos \phi + b \right)^2 + \left(\ell _{\rm max} \sin \phi \right)^2 = 4 R^2_A \ .
\label{34}
\end{eqnarray}
Thus with the definition $\hat{\ell} = \ell _{\rm max}/2 R_A$ we have 
\begin{eqnarray}
\hat{\ell} (b, \phi) = - \hat{b}  \cos \phi + ( 1 - \hat{b}^2  \sin ^2 \phi )^{1/2} \ .
\label{35}
\end{eqnarray}
This is a normalized path length that quantifies the dependence on $b$ and $\phi$.  Assuming that the energy loss is proportional to the square root of the initiating parton momentum \cite{rb,rf}, we write $\Delta k$ in the form
\begin{eqnarray}
\Delta k =  \epsilon (b)  \hat{\ell} (b, \phi) \sqrt{k'} \ ,
\label{36}
\end{eqnarray}
where $\epsilon (b)$ is the energy-loss coefficient that may be given a reasonable form \cite{rf}
\begin{eqnarray}
 \epsilon (b) = \epsilon_0  {1 - e^{-2 (1-\hat{b})}\over 1 - e^{-2}}                 \ ,
\label{37}
\end{eqnarray}
since the density decreases with increasing $b$.  The coefficient $\epsilon _0$ is to be determined below.

In Ref.\  \cite{rh} we have found that the shower parton distribution at mid-rapidity in a central heavy-ion collision is given by
\begin{eqnarray}
{\cal S}_j (q) = \xi \sum_i \int dk k f_i (k) S^j_i(q/k) \ ,
\label{38}
\end{eqnarray}
where $f_i (k)$ is the distribution of hard partons $i$ without nuclear suppression and $\xi = 0.07$ is the suppression factor that is necessary to fit the normalization of the pion spectrum at intermediate $p_T$ by TS recombination and is in accord with $R_{AA}$ being $\sim 0.2$ at higher $p_T$.  We must now generalize Eq.\ (\ref{38}) to non-central collision.  When $b>0$, there is $\phi$ dependence in $\Delta k$, given by Eq.\ (\ref{36}).  Instead of the suppression factor $\xi$, we can describe the energy loss of the hard parton by shifting the parton momentum and writing ${\cal S}_j(q)$ as
\begin{eqnarray}
\xi \sum_i \int dk k f_i (k) S^j_i\left({q \over k}\right) \hskip3cm\\  \nonumber
 = \sum_i \int dk k f_i (k + \Delta k) S^j_i\left({q \over k}\right)
\label{39}
\end{eqnarray}
at $b = 0$, and then generalize the RHS to $b >0$ and endowing it with a $\phi$ dependence through Eqs.\  (\ref{35}) and (\ref{36}).  We note that $k$ is the momentum of the hard parton that emerges from the medium, and $q$ is the momentum of the parton $j$ in the shower, so $k$ is to be integrated from the lower limit at $q$.  The momentum $k' = k + \Delta k$ is that of parton $i$ immediately after hard scattering and before traversing the medium.

Writing Eq.\ (\ref{36}) as $k' = k+ \epsilon \hat{\ell} \sqrt{k'}$, we can solve for $k'(k, b, \phi)$, which, when substituted into $f_i(k')$, yields a function of $k$ that is denoted by $F_i (k, b, \phi)$ in Eq.\ (\ref{33}), i.e.,
\begin{eqnarray}
 f_i \left(k' (k, b, \phi)\right) = 2\pi F _i (k, b, \phi) , 
 \label{40}
\end{eqnarray}
where $F_i$ is defined per radian, while $f_i (k') $ is integrated over all $\phi$.  The distribution $f_i (k')$ has a power-law behavior 
\begin{eqnarray}
 f_i (k' ) = {a \over \left(1 + k'/k_0 \right)^{\beta}}\ , 
 \label{41}
\end{eqnarray}
where the parameter $a$, $k_0$ and $\beta$ are given in Ref.\ \cite{ds} for a variety of parton type $i$.  Thus $F _i (k, b, \phi)$ can be written in the form
\begin{eqnarray}
F _i (k, b, \phi)  = {1 \over 2\pi}f_i (k)  G(k, b, \phi) \ ,
\label{42}
\end{eqnarray}
where
\begin{eqnarray}
G(k, b, \phi) = \left[ 1 +  {\Delta k (k, b, \phi)  \over k + k_0}  \right]^{- \beta}  \ .
\label{43}
\end{eqnarray}
Keeping only the zeroth and second harmonics of the $\phi$ dependence, we have
\begin{eqnarray}
G(k, b, \phi) = g_0(k, b) + 2 g_2(k, b) \cos 2 \phi ,
\label{44}
\end{eqnarray}
where $g_0$ and $g_2$ can be determined explicitly in terms of $\epsilon_0$ and $\hat{b}$, beside $k$, as we shall show in subsection {\bf C} below.

\subsection{ TS+SS Recombination}

Having obtained the $\phi$ dependence of the hard parton distribution per radian at the medium surface, $F_i(k, b, \phi)$, we can proceed to the shower partons for recombination with the thermal partons.  We have
\begin{eqnarray}
{\cal S}_j (q_1, b, \phi) = \sum_i \int dk k F_i(k, b, \phi) S^j_i(q_1/k) \ , 
\label{45}
\end{eqnarray}
\begin{eqnarray}
{\cal T}_{j'} (q_2, b, \phi) = C(b) q_2 e^{-q_2/\tilde T'(b, \phi)} \ , 
\label{46}
\end{eqnarray}
\begin{eqnarray}
{dN^{\rm TS}_{\pi}(b) \over p_Tdp_Td\phi} = {1 \over p^3_T} \int^{p_T}_0 dq_2\ {\cal T}(q_2,  b, \phi){\cal S}(p_T-q_2, b, \phi) ,
\label{47}
\end{eqnarray}
where the quark types $j$ and $j'$ are paired in appropriate ways to form a pion of a specific charge; for example for $\pi^+$, they are $u\bar{d}$ and $\bar{d}u$.  The pion recombination function has already been taken into account in the derivation of Eq.\ (\ref{47}) \cite{rh}.  The values of $C(b)$ are given in Ref.\ \cite{rh8}.  The thermal parton \dis\ ${\cal T}(q, b, \phi)$ is $B+R$ with a new inverse slope $\tilde T'(b, \phi)$ that depends on $b$ and $\phi$ because it is a measure of the enhanced thermal partons in response to the hard parton with momentum $k$ and angle $\phi$. Equation (\ref{47}) should not be considered for $p_T < 2$ GeV/c, since the validity of the formalism is questionable at low $p_T$ and TS is dominated by TT recombination anyway.

Unlike the $\phi$ dependence at low \pt\ where semi-hard partons are restricted to $\phi\in\cal R$, now the hard parton can originate from the interior of the medium and can be directed at any $\phi$. The thermal partons in the vicinity of the trajectory are enhanced in proportion to the local medium density which is dependent on $b$ and $\phi$. Since that density at fixed $0<\hat b<1$ decreases with increasing $\phi$,  we adopt the reasonable form 
\begin{eqnarray}
\tilde T'(b, \phi) =  T' \left({1+a\hat b \cos\phi\over 1+2a\hat b/\pi}\right) ,   \label{47a}
\end{eqnarray}
whose average over $0<\phi<\pi/2$ is the value $T'=0.3$ GeV used in Sec.\ III.A. The size of the parameter $a$ awaits revelation by data on the ridge dependence on the trigger azimuthal angle, about which there is preliminary supportive indication from STAR. We shall use the provisional value $a=0.1$ to carry out the calculation in Sec.\ IV.D.

For SS recombination it is important to recognize that only one hard parton is involved and that the two shower partons are from the same jet.  Thus the formalism is the same as described in Ref.\ \cite{rh}, the only difference being the replacement of $\xi f_i (k)/2 \pi$ by $F_i (k, b, \phi)$ for non-central collision, i.e., 
\begin{eqnarray}
{dN^{\rm SS}_{\pi}(b) \over p_Tdp_Td\phi}= {1 \over p^3_T} \sum_i \int dk k F_i (k, b, \phi) \nonumber\\ 
\times \int^{p_T}_0 dq \{S, S\} (q, k, p_T)  \ ,
\label{48}
\end{eqnarray}
where the curly brackets denote the symmetrization of the leading parton momentum fractions $z_1=q/k$ and $z_2=(p_T-q)/k$
\begin{eqnarray}
 \{S, S\} (q, k, p_T) = {1\over 2}\left[S^j_i\left(z_1\right)S^{j'}_i\left({z_2 \over 1-z_1}\right) \right.\nonumber \\ 
 + \left.S^j_i\left({z_1\over 1-z_2}\right)S^{j'}_i\left(z_2  \right)\right] \ .
\label{49}
\end{eqnarray}
As explained in Ref.\ \cite{rh}, the last integral in  Eq.\ (\ref{48}) is essentially the fragmentation function $D(p_T/k)$, so the equation describes both recombination and fragmentation, neither of which has $\phi$ dependence.  Clearly, QNS cannot be valid when the only $\phi$ dependence arises from the hard parton distribution $F_i (k, b, \phi)$.   Let us write the sum of Eqs.\ (\ref{47}) and (\ref{48}) symbolically as 
\begin{eqnarray}
{dN^{\rm sh}_{\pi}(b) \over p_Tdp_Td\phi} =  F_i (k, b, \phi) \otimes  ({\cal T} S + SS)  \ ,
\label{50}
\end{eqnarray}
where the superscript sh denotes distributions involving shower partons.

\subsection{Momentum Shift}

Before proceeding to the calculation of $v^{\pi}_2 (p_T, b)$, let us first determine the normalization of the inclusive distribution averaged over all $\phi$.  It follows from Eqs.\ (\ref{42}), (\ref{44})  and (\ref{50}) that
\begin{eqnarray}
{dN^{\rm sh}_{\pi}(b) \over p_Tdp_T}&=& {1 \over 2\pi} \int^{2\pi}_0 d\phi {dN^{\rm sh}_{\pi}(b) \over p_Tdp_Td\phi}   \nonumber\\
&=& {1 \over 2\pi} g_0 (k, b)    f_i (k) \otimes ({\cal T} S + SS)  \ . \label{51}
\end{eqnarray}
where the  $\phi$ dependence in $\cal T$ is neglected, since this equation will be used only for 0-10\% centrality in this subsection.
For the TT component we take the average of $B$ and $B+R$, and get
\begin{eqnarray}
{dN^{\rm th}_{\pi}(b) \over p_Tdp_T} = {2 \over \pi} \left\{ \left({\pi \over 2} - \Phi \right) B_{\pi}(p_T) + \Phi \left[B_{\pi}(p_T)  + R_{\pi}(p_T) \right]\right\} \nonumber\\
= {C^2(b) \over 3 \pi} \left\{ \left[{\pi \over 2} - \Phi (b) \right]e^{-E_T(p_T)/T} + \Phi (b) e^{-E_T(p_T)/T'} \right\}. 
\label{52}
\end{eqnarray}

\begin{figure}[tbph]
\includegraphics[width=0.45\textwidth]{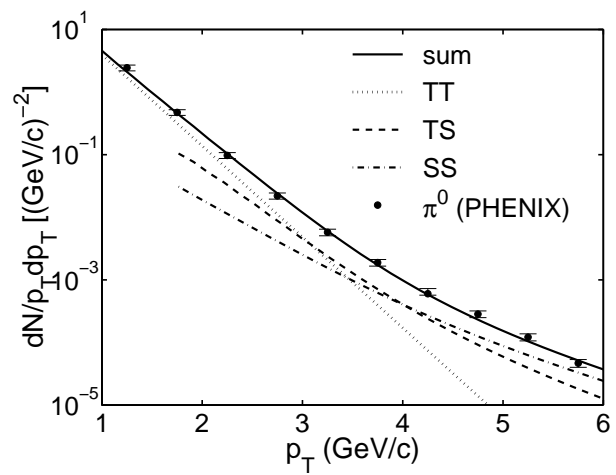}
\caption{Fit of the $\pi^0$ distribution at one point, $p_T=0.435$ GeV/c, by adjusting the energy loss coefficient $\epsilon_0$. The data are for 0-10\% centrality from Ref.\ \cite{sa2}.}
\end{figure}

The sum of these two equations gives the total inclusive pion distribution $dN_{\pi}(b)/p_T dp_T$.  It is to be compared with the data, labeled $dN/2\pi p_Tdp_T$, where $N$ refers to the number of pions per event  integrated over all $\phi$.  We have one parameter in Eq.\ (\ref{51}), which is $\epsilon_0$ in Eq.\ (\ref{37}).  It determines the scale of momentum shift due to energy loss of the hard parton in the medium, and enters Eq.\ (\ref{51}) through $g_0(k, b)$.  The larger $\epsilon_0$ is, the smaller is $g_0(k, b)$ and the more suppressed is the shower parton contribution.  We determine $\epsilon_0$ by fitting the normalization of the data for 0-10\% centrality.  Note that the $p_T$ dependence is not adjustable, but the overall degree of suppression of the shower component is adjustable, since $\epsilon_0$ is not determined within our formalism. Fitting the normalization at just one point ($p_T=4.35$ GeV/c) we obtain
\begin{eqnarray}
\epsilon_0 = 0.55 \,  {\rm GeV}^{1/2} \ .
\label{53}
\end{eqnarray}
The rest of the $p_T$ distribution is shown by the solid line in Fig.\ 6 in excellent agreement with the data \cite{sa2} for $\sqrt {s_{NN}}=200$ GeV, the same set as used in \cite{rh}.

\begin{figure}[tbph]
\includegraphics[width=0.45\textwidth]{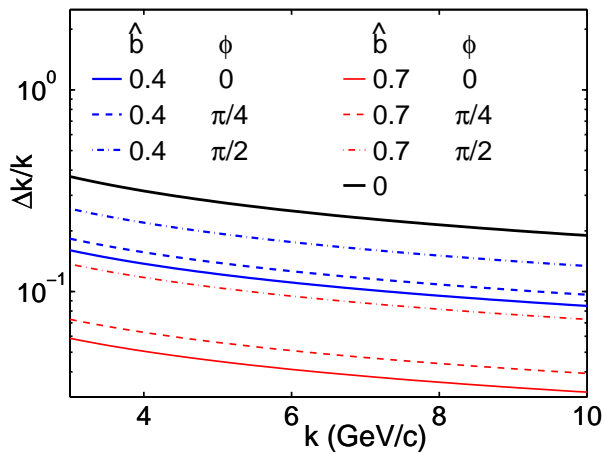}
\caption{(Color online) Factional momentum shift for various normalized impact  parameter $\hat b=b/2R_A$ and $\phi$.}
\end{figure}

Since we have obtained $k'(k, b, \phi)$ in terms of $\epsilon \hat{\ell}$ already, we can determine the fractional shift $\Delta k/k$, shown in Fig.\ 7, for three values of normalized impact parameter, $\hat{b} = 0$, 0.4, and 0.7, and three values of $\phi$ at 0, $\pi/4$, and $\pi/2$.  They all decrease with $k$ roughly as $1/\sqrt{k}$, as expected.  There is, of course, no dependence on $\phi$ when $\hat{b} = 0$.  As $\hat{b}$ increases, the nuclear overlap region gets smaller, and $\Delta k/k$ becomes smaller.  At fixed $\hat{b} > 0$,  $\Delta k/k$ is smaller at $\phi = 0$ (in solid lines) than at $\phi = \pi/2$ (in dashed-dot lines) because the average path length is shorter.  Since $\beta$ in Eq.\ (\ref{43}) is around 8, $G(k, b, \phi)$ can be quite small, if $\Delta k/k$ is not infinitesimal.  We show in Fig.\ 8 the harmonic components  $g_0(k, b)$ and $g_2(k, b)$ defined in Eq.\ (\ref{44}) for $\hat{b} = 0$, 0.4 and 0.7. The $\phi$ dependence of the hard parton is important because it gives the dominant contribution to the hadronic $v_2$.  As $\hat{b}$ increases $g_0(k, b)$ increases rapidly, indicating less suppression.  For the lines showing $g_2(k, b)$, which are amplified by a factor of 10 in Fig.\ 8, we see that their values are between 0.03 and 0.04, relatively insensitive to $k$, and, of course, $g_2(k, 0) = 0$.  It is the ratio $g_2(k, b)/g_0(k, b)$ that sets the scale for $v_2$ of hard partons.

\begin{figure}[tbph]
\includegraphics[width=0.45\textwidth]{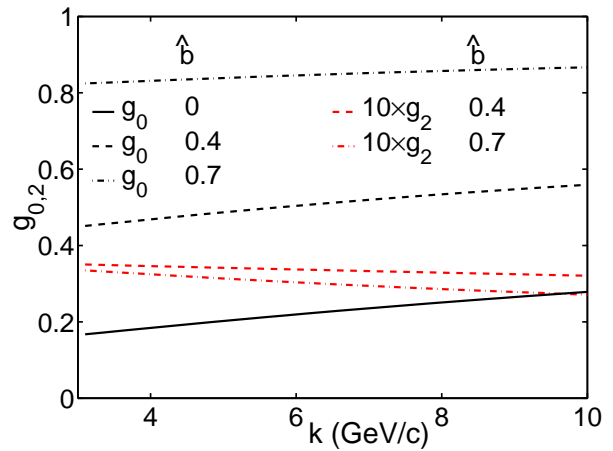}
\caption{(Color online) The functions $g_0(k,b)$ and $g_2(k,b)$ for some typical values of $\hat b$.}
\end{figure}

\subsection{$\bf v_2$ for Pion}

We are now ready to calculate $v_2^{\pi}(p_T, b)$.  
For the contribution from the shower partons at intermediate \pt\ we write out Eq.\ (\ref{50}) fully as
\begin{eqnarray}
{dN_{\pi}^{\rm sh}\over p_Tdp_Td\phi}&=& {1 \over  p^3_T} \sum_i \int^{\infty}_3 dk k F_i(k, b, \phi) \nonumber \\
&&\times \int^{p_T}_0 dq\ \left[C(b)qe^{-q/\tilde T'(b,\phi)}S^j_i \left({p_T - q\over k} \right)\right. \nonumber \\
&&+ \left. \left\{S^j_i \left({q\over k}\right), S^{j'}_i \left({p_T - q\over k - q}\right) \right\} \right] \ .
\label{55}
\end{eqnarray}
Putting this in Eq.\ (\ref{6}) and using (\ref{42}), (\ref{44}) and (\ref{47a}) for the $\phi$ dependences, we perform the integrations over $\phi$ first, and then over $q$ and $k$. The result is $v_2^{\pi}(p_T,b)$ for $p_T>3$ GeV/c. That is shown in  Fig.\ 9 for the high $E_T$ portion in that figure. We note that the effect of $\phi$ dependence of $\tilde T'(b, \phi)$ on the magnitude of $v_2^{\pi, {\rm sh}}(p_T, b)$ in the high $E_T$ region is only about 10\%. Thus the main contribution to $v_2^{\pi, {\rm sh}}(p_T, b)$ comes from the $\phi$ dependence of $F_i(k, b, \phi)$ for the hard parton in Eq.\ (\ref{55}), not from that of the thermal parton $\cal T$ in (\ref{50}). If we neglected the  $\phi$ dependence of $\tilde T'(b, \phi)$ in (\ref{55}), we would have a simple  formula in closed form
\begin{eqnarray}
v^{\pi, {\rm sh}}_2 (p_T, b) = {g_2(k, b) f_i (k) \otimes \left({\cal T}S + SS \right) \over g_0(k, b) f_i (k) \otimes \left({\cal T}S + SS \right)} ,
\label{55a}
\end{eqnarray}
which offers a succinct exhibit of all the important factors for the main contribution to $v_2$ from the hard parton  at high $E_T$.

\begin{figure}[tbph]
\includegraphics[width=0.45\textwidth]{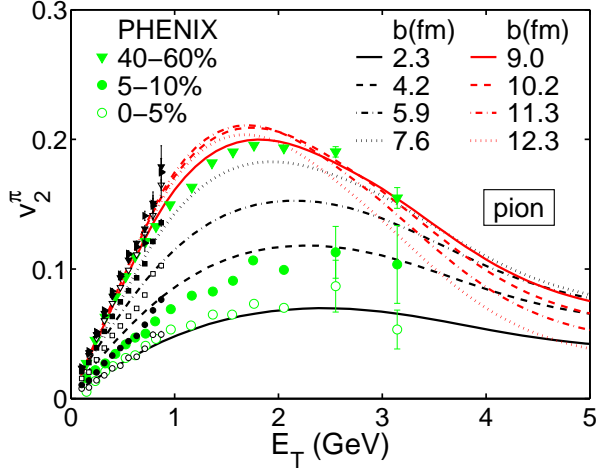}
\caption{(Color online) $v_2^{\pi}$ for a wide range of $E_T$. The small symbols are the same as those in Fig.\ 1; the larger symbols (in green) are preliminary data from \cite{phen}.}
\end{figure}

For the low \pt\ part the contribution from thermal partons is what we obtained in Sec.\ III.A, which we can write out explicitly from Eqs.\ (\ref{7}) and (\ref{20})
\begin{eqnarray}
v^{\pi, {\rm th}}_2 (p_T, b) = {\sin 2 \Phi (b) \over \pi /\left[ e^{E_T (p_T)/T''_{\pi}-1}\right] + 2 \Phi (b)} \ .
\label{56}
\end{eqnarray}
The result has been shown in Fig.\ 1, and is shown again in Fig.\ 9 for $E_T<1$ GeV. 

The high and low $E_T$ regions should be connected by an interpolating function that spans the two regions.  That function should depend on the relative weight of the $p_T$ distributions given in Eqs.\ (\ref{51}) and (\ref{52}).  For notational brevity let us use {\bf TT} to denote $dN^{\rm th}_\pi (b)/p_Tdp_T$, and {\bf TS} + {\bf SS} to denote $dN^{\rm sh}_\pi (b)/p_Tdp_T$, which   is the average of Eq.\ (\ref{55}) over $2\pi$ of $\phi$. During the interpolation procedure
that equation should not be used for $p_T$ too small.  To ensure that, we cut it off by hand by inserting a cut-off function,  $1 - \exp (-0.1 p^3_T)$, which decreases from $\sim 1$ to $\sim 0$ as $p_T$ is decreased from 3 to 1 GeV/c.  This damping factor has been used in Fig.\ 6.  We construct the weight function
\begin{eqnarray}
W (p_T, b) = {{\bf TT} \over {\bf TT}+{\bf TS} + {\bf SS}} \ ,
\label{57}
\end{eqnarray}
in terms of which we define the overall $v_2^{\pi}$ by
\begin{eqnarray}
 v ^{\pi}_2 (p_T, b) &=& v ^{\pi, \rm th}_2 (p_T, b) W(p_T, b) \nonumber \\
 && +v ^{\pi, \rm sh}_2 (p_T, b) \left[1 - W(p_T, b)\right] \ .
\label{58}
\end{eqnarray}
Our calculated results for $ v ^{\pi}_2 (p_T, b)$ are shown in Fig.\ 9 in terms of $E_T$ for $0 < E_T< 5$ GeV at various values of $b$ in fm, chosen to correspond to the centralities 0-5\%, 5-10\%, 10-20\%, 20-30\%, etc., as in Fig.\ 1, with $\hat{c}$ being approximately $\hat{b}^2$.  It is evident that in all cases $v ^{\pi}_2$ increases at low $E_T$ and decreases at high $E_T$, reflecting the different mechanisms responsible for $\phi$ anisotropy.

The data for $E_T < 1$ GeV in small symbols are from Ref.\ \cite{ja4} already shown in Fig.\ 1.  For $E_T > 1$ GeV we show the data from Ref.\ \cite{phen} for 3 centrality bins:  0-5\%, 5-10\%, 40-60\%; they are indicated by large symbols.  The first ones (0-5\%) in large open circles correspond to $b = 2.3$ fm; the next (5-10\%) in large filled circles correspond to $b= 4.2$ fm.  The agreement with our results are good.  There are no more centrality bins in Ref.\ \cite{phen} that correspond to what we have calculated.  We show only 40-60\% in large triangles, which correspond to $b = 10.2$-11.3 fm among our calculated curves.   
On the whole our result reproduces the data quite well.

\subsection{$\bf v_2$ for Proton}

For proton production we consider TTT, TTS and TSS recombination, leaving out SSS which is not important unless $p_T > 8$ GeV/c.  The complication of 3-quark recombination has been treated in Ref.\ \cite{rh} already; it does not affect the calculation of $v _2$, since hadronization occurs after jet quenching as we have seen in the pion case.  Thus we proceed as before using Eq.\ (\ref{7}) for $v^{\pi, {\rm th}}_2 (p_T, b)$ for $p_T<1$ GeV/c with appropriate $B_p(p_T, b) / R_p(p_T, b)$ discussed in Sec. III.B.  At intermediate $p_T$ we generalize Eq.\ (\ref{50}) by including one more thermal parton, i.e.,
\begin{eqnarray}
{dN^{\rm sh}_{p}(b) \over p_Tdp_Td\phi} =  F_i (k, b, \phi) \otimes  ({\cal T}{\cal T} S + {\cal T}SS)  \ .
\label{59}
\end{eqnarray}
When written out in full, it looks like Eq.\ (\ref{55}) but with an extra factor, ${\cal T}(q_1, b, \phi)$, attached to each of the two terms in (\ref{55}). For those thermal partons we use Eqs.\ (\ref{46}) and (\ref{47a}), except that $T'$ in (\ref{47a}) has a $b$ dependence, given in (\ref{29}).
The  calculation of $v_2^{p,{\rm sh}}(p_T, b)$ can then be carried out as before, but with an extra integration over $q_1$. 
Although there is no analytical formula for the result of that calculation, a simplified form that neglects the $\phi$ dependence of $\tilde T'(b, \phi)$ is
\begin{eqnarray}
 v ^{p, \rm sh}_2 (p_T, b) = {g_2 (k, b) f_i (k) \otimes \left({\cal T} {\cal T}S +  {\cal T}SS \right)\over g_0 (k, b) f_i (k) \otimes \left({\cal T} {\cal T}S +  {\cal T}SS \right)} \ ,
\label{59a}
\end{eqnarray}
which is the counter part of Eq.\ (\ref{55a}). The quantitative results shown below is, however, from the full calculation.

The overall $v_2^p(p_T, b)$ for all \pt\ region again involves a weighted average
\begin{eqnarray}
 v ^p_2 (p_T, b) &=& v ^{p,{\rm th}}_2 (p_T, b) W(p_T, b) \nonumber \\
 &&+v ^{p,{\rm sh}}_2 (p_T, b) \left[1 - W(p_T, b)\right] \ .
\label{60}
\end{eqnarray}
where the corresponding weight function is
\begin{eqnarray}
W (p_T, b) = {{\bf TTT} \over {\bf TTT}+{\bf TTS} + {\bf TSS}} \ .
\label{61}
\end{eqnarray}
For {\bf TTT} we have
\begin{eqnarray}
{dN^{\rm th}_p \over p_Tdp_T} = {2Ap^2_T \over \pi m_T} \left[\left({\pi \over 2} - \Phi  \right)e^{-E_T(p_T)/T}  + \Phi e^{-E_T(p_T)/T'}\right],
\label{62}
\end{eqnarray}
while for {\bf TTS} and  {\bf TSS} we have 
\begin{eqnarray}
{dN^{\rm sh}_p \over p_Tdp_T}& &= {1 \over 2\pi} \sum_i  \int dk kf_i(k)g_0(k) \nonumber \\
&& \times \int dq_1 dq_2 \left({\cal T} {\cal T} S + {\cal T} SS\right) \ .
\label{63}
\end{eqnarray}
where the average ${\cal T}$ over $\phi$ is used, which means setting $\tilde T'(b,\phi)=T'(b)$.

\begin{figure}[tbph]
\includegraphics[width=0.45\textwidth]{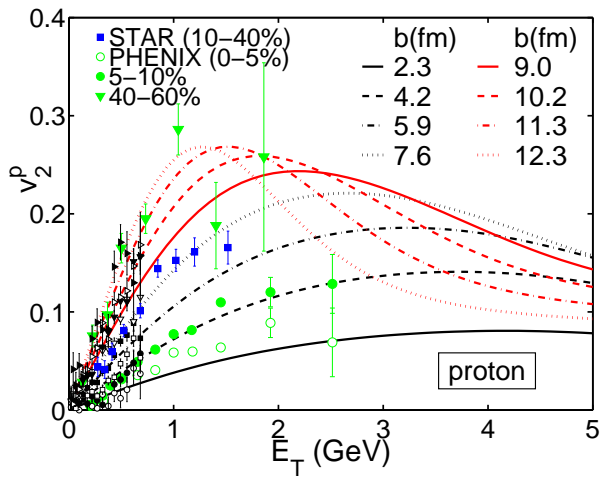}
\caption{(Color online) $v_2^{p}$ for a wide range of $E_T$. The small symbols are the same as those in Fig.\ 4. The larger symbols (in green) are preliminary PHENIX data \cite{phen} for centralities  0-5\% (open circle), 5-10\% (full circle), and 40-60\% (triangle). The STAR data are for 10-40\% (blue square) at $\sqrt {s_{NN}}=62.4$ GeV \cite{ba}.}
\end{figure}

The result on $v^p_2 (E_T, b)$ is shown in Fig.\ 10.  The curves exhibit similar patterns as those for $v^{\pi}_2 (E_T, b)$, but are generally higher, and persist to rise at high $b$ when $E_T<1.5$ GeV.  Although the $\phi$ dependence of the hard parton is the same for $\pi$ or $p$, the thermal partons are not the same for the two cases.  Since two thermal partons can participate in the formation of proton, the $\phi$ dependence of $\tilde T'(b, \phi)$ plays a larger role in $v_2^{p,{\rm sh}}(p_T, b)$ than in $v_2^{\pi,{\rm sh}}(p_T, b)$.
  The data in the higher $E_T$ region are shown by the same symbols as in Fig.\ 9.  For 0-5 \% and 5-10\% centralities the data are from Ref.\ \cite{phen} for $p + \bar p$.  However, for 40-60\% centrality the data in \cite{phen} for $p + \bar{p}$ differ significantly from the data in \cite{sa3} for $p$ alone.  It is the latter in large triangles that we show in Fig.\ 10. We include in that figure also the STAR data \cite{ba} at 62.4 GeV for $p+\bar p$ at 10-40\% centrality that agree well with our dotted line for 20-30\%. 
   For all the centralities shown the data are in general accord with our results.
   
To summarize this long section we point out first that there is only one free parameter $\epsilon_0$ used to determine the medium suppression of hard parton by fitting the pion \dis\ at one point, $p_T=4.35$ GeV/c in Fig.\ 6. The parameter $\epsilon_0$ replaces the suppression factor $\xi$ used earlier \cite{rh} for central collision only without considering explicitly the issue of energy loss. The shape of the \pt\ \dis\ in Fig.\ 6 is calculated. The path length dependence of jet quenching introduces the $\phi$ dependence at high \pt\ that is the main source of elliptic flow.
The characteristics of $v_2^h(E_T,b)$ for $h=\pi$ and $p$ in Figs.\ 9 and 10 are determined without any more adjustable parameter. Although there are some minor discrepancies between our results and the $v_2$ data, on the whole we have been able to reproduce the data very well.

\section{Breaking of quark number scaling}

In the naive application of the recombination model there is quark number scaling (QNS) of $v_2$, which may be stated as the universality of $v^h_2 (p_T/n_q)/n_q$ where $n_q$ is the number of constituent quarks in the hadron $h$ \cite{dm,pk3}.  The simple argument used is based on the factorizability of the distribution of the quarks that recombine, and on the simplification of the recombination function to the form that contains $\delta (q_j - p_T/n_q)$ for each of those quarks.  There are also other considerations for the origin of QNS \cite{pp}.
Experimental verification of QNS has evolved to the replacement of $p_T$ by $E_T$ with impressive confirmation of the scaling behavior \cite{ba}-\cite{aa}, at least at low $E_T/n_q$.  Since it is known that fragmentation is more important than recombination at very high $p_T$ (or in very peripheral collisions), QNS should break down at some point \cite{vg,rf}.  The question is at what point.  We show here that it occurs rather early, even when TS recombination is still dominant.  In fact, at even lower $p_T$ where TT and TTT recombination are more important, QNS is not valid in general for specific centralities, as discussed in Sec.\ III.B.  However, averaging over all centralities leads to approximate QNS, in agreement with minimum bias data \cite{ba,at,ps,aa}.

The breaking of QNS is due mainly to the breaking of factorization of joint parton distribution, which, if true, would have the form
\begin{eqnarray}
F_{n_q}\left(q_1, \phi_1;\cdots; q_{n_q},\phi_{n_q}\right)=\prod^{n_q}_{i = 1} F_i (q_i,\phi_i) \nonumber \\
=\prod^{n_q}_{i = 1} F_i (q_i) \left(1 + 2v^i_2 (q_i) \cos 2 \phi _i \right).
\label{64}
\end{eqnarray}
From this follows
\begin{eqnarray}
v^h_2 (p_T) \simeq n_q v^q_2 (p_T/n_q)
\label{65}
\end{eqnarray}
for $q_i = p_T/n_q$.  For pion at low $p_T$ the factorization of $q\bar{q}$ distribution is reasonable, and that is how Eqs.\ (\ref{11}) and (\ref{12}) are obtained at all centralities.  For proton at low $p_T$ the factorization of $uud$ distribution becomes questionable for non-central and peripheral collisions, the consequence of which is that the inverse slope for proton differs from that of the single inclusive $u$ (or $d$) quark.  Instead of investigating the non-factoriable form of $F_{uud}(q_1, q_2, q_3)$, we have adopted in Sec.\ III.B  phenomenological form for the average of $B_p(p_T) + R_p (p_T,\phi)$ over all $\phi$ with $T'_p$ given in Eq.\ (\ref{29}) and shown in Fig.\ 3.  That results in the centrality dependence of $T''_p$ given in Eq.\ (\ref{31}).  Since $T''_{\pi}$ is constant for all $c$ at the value specified in Eq.\ (\ref{18}), $v^h_2 (p_T,b)$, as given by Eq.\ (\ref{8}), has initial $E_T$ dependences that are different for pion and proton, as expressed explicitly in Eqs.\ (\ref{22}) and  (\ref{32}), except at one point of cross-over that occurs at a centrality roughly equivalent to minimum bias.

At intermediate $p_T$ where the contribution from shower partons is important, the $\phi$ dependence comes mainly from $F_i (k, b, \phi)$ for the hard parton in Eq.\ (\ref{50}) for pion and Eq.\ (\ref{59}) for proton.  Thus the corresponding  $v^h_2 (p_T,b)$ for $h = \pi$ and $p$ are given primarily by Eqs.\ (\ref{55a}) and (\ref{59a}) that have the same structure, the common factors of importance being $g_2 (k,b)$ in the numerators and $g_0 (k, b)$ in the denominators.   The significance of $g_2 (k,b)/g_0 (k, b)$ being the major factor for both $v^{\pi}_2$ and $v^p_2$ immediately implies that 
\begin{eqnarray}
{1 \over 2}v^{\pi}_2 (E_T/2, b) > {1 \over 3} v^p_2 (E_T/3, b)
\label{66}
\end{eqnarray}
at intermediate $E_T$ for all $b$.  However, when the $\phi$ dependence in $\cal T$ in those two equations is taken into account, the results on $v_2^{h,{\rm sh}}$ are enhanced more for proton than for pion. That makes the inequality in Eq.\ (\ref{66}) less unbalanced. Nevertheless, QNS is broken because not all partons contribute equally to the azimuthal anisotropy, even though both hadrons are formed by recombination.

Since Figs.\ 9 and 10 exhibit all properties of $v^{\pi}_2 (E_T, b) $ and $v^p_2 (E_T, b)$ that we have obtained, we can calculate $v^{\pi}_2 (E_T/2, b)/2 $ and $v^p_2 (E_T/3, b)/3$ and show their differences for four centralities in Fig.\ 11.  Although the pion (dashed) and proton (dashed-dotted) lines start out with similar slopes at low $E_T$, they separate at higher $E_T$, the former being consistently higher than the latter.  The data that support QNS are for minimum bias and at low $E_T$ where the theoretical lines are in agreement with the data at all four centralities.  The breaking of QNS becomes visibly clear at $E_T/n_q > 0.5$ GeV, and is discernible even in the minimum bias data of Ref.\ \cite{ba}.

\begin{figure}[tbph]
\includegraphics[width=0.45\textwidth]{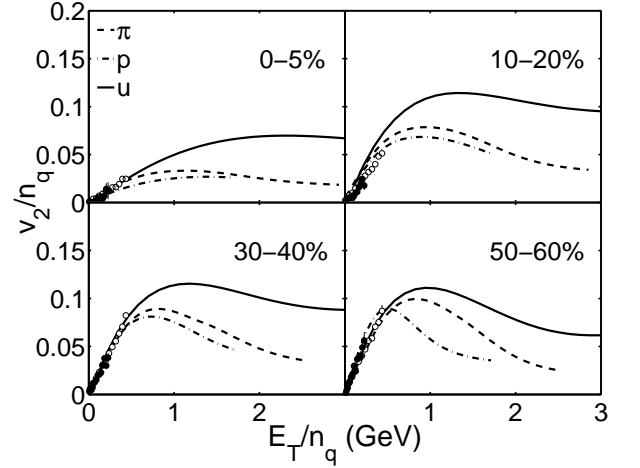}
\caption{Scaled $v_2$ for pion, proton and $u$ quark. For the $u$ quark the thermal $v_2$ is plotted at low $E_T$, changing to shower $v_2$ at high $E_T$.
The data points are scaled from those in Fig.\ 1 (open circles) and Fig.\ 4 (filled circles) from \cite{ja4}. At higher $E_T/n_q$ only minimum bias data are available, not suitable for display here.}
\end{figure}

The reasons for plotting $v^h_2 (E_T/n_q)/n_q$ are not only to check QNS, but also to give a hint of the elliptic flow of the recombining quarks, if QNS were valid.  We can calculate $v^q_2 (E_T, b)$ for quarks theoretically before recombination.  At low $E_T$ Eq.\ (\ref{7}) is valid for quarks also, with $B_q(q)/R_q(q)$ being given by Eq.\ (\ref{20}).  For  $E_T$ we use the constituent quark mass, which we take to be 0.3 GeV.  The resulting $v^q_2 (E_T, b)$ may be regarded as reliable for $E_T\ ^<_\sim\  0.7$ GeV. For higher $E_T$ we recall Eqs.\ (\ref{50}) and (\ref{55a})  for pions.  The structure for quarks is similar, except that there is no recombination.  Thus for a quark of type $j$ in the shower we have
\begin{eqnarray}
{dN^{\rm sh}_j (b) \over q dq d \phi} = \sum_i \int dk k F_i (k, b, \phi) S^j_i (q/k) \ ,
\label{67}
\end{eqnarray}
\begin{eqnarray}
v^{j,{\rm sh}}_2 (q, b) = {\int dk k g_2 (k, b) \sum_i f_i(k) S^j_i (q/k) \over \int dk k g_0 (k, b) \sum_i f_i(k) S^j_i (q/k)}\ .
\label{68}
\end{eqnarray}
These equations are for shower partons with momentum $q$ not too low.  The lower bound for their validity may be set at $q \sim$ 1.5-2.0 GeV/c.  For illustrative purpose we consider $u$ quark for definiteness.  For each centrality in Fig.\ 11 we show $v^u_2 (q, b)$ for the two regions: thermal quark for $E_T < 0.7$ and shower quark for $E_T> 1.5$ GeV; in between we simply connect the two by a smooth interpolating curve without mathematical significance.  The physical significance of  $v^u_2 (E_T, b)$ shown is that it is arising function of $E_T$ at low  $E_T$ and a gently falling function at higher $E_T$, characterizing the thermal and shower partons, respectively.

It is now evident from Fig.\ 11 that the three curves for $u$, $\pi$ and $p$ for each centrality are distinctly different. From central to mid-central collisions up to 40\% centrality the difference between $\pi$ and $p$ curves may not be large enough for the present data on $v_2$ to discriminate.
There is some evidence for QNS breaking at $E_T/n_q>0.5$ GeV in the minimum bias data  for $\pi$ and $p$ \cite{yb}. They are for $\sqrt {s_{NN}}=62.4$ GeV. Roughly, the pion data are above the proton data, as in Fig.\ 11.
Current experimental efforts have largely been to make corrections due to fluctuations in eccentricity in order to achieve QNS. Our suggestion is to focus on individual centrality bins and quantify the breaking of QNS.

The results shown in Fig.\ 11 are from extensive and detailed calculation. It would be helpful if the quantitative difference between the two $v_2^h$ behaviors for pion and proton can be explained in simple terms, albeit inexact and schematic. To that end let us assume dominance of TS and TTS recombination for $E_T/n_q\ ^>_\sim\ 1$ GeV and write
\begin{eqnarray}
v_2^{\pi}=\left<\cos 2\phi\right>_{\rm TS} \ , \qquad v_2^{p}=\left<\cos 2\phi\right>_{\rm TTS} \ .  \label{71}
\end{eqnarray}
If the contribution to $v_2$ from thermal and shower partons are expressed as $v_2^{\rm T}$ and $v_2^{\rm S}$, respectively, then  we have approximately
\begin{eqnarray}
v_2^{\pi}\simeq v_2^{\rm T}+v_2^{\rm S}\ , \qquad 
v_2^{p}\simeq 2v_2^{\rm T}+v_2^{\rm S}\ ,  \label{72}
\end{eqnarray}
where the momenta of the quarks are not shown explicitly because they are complicated to express. Unlike the naive description given in Eq.\ ({67}), the thermal parton momentum is smaller compared to the shower parton momentum because the \dis\ of the former is damped exponentially while that of the latter is power suppressed. Furthermore, even if the quark momentum is $q=p_T/n_q$, we have $v_2^{\rm T}(q)<v_2^{\rm S}(q)$ because of the difference in the $\phi$ dependence in T and S. That inequality becomes more unequal when smaller $q$ is used in $v_2^{\rm T}(q)$ and larger $q$ in $v_2^{\rm S}(q)$.  Thus roughly we have at some fixed $E_T>2$ GeV
\begin{eqnarray}
{v_2^{\pi}(E_T)\over v_2^{p}(3E_T/2)}\simeq {2+\delta\over 3+\delta}>{2\over 3}\ , \label{73}
\end{eqnarray}
where
\begin{eqnarray}
\delta=v_2^{\rm S}(q_+)/v_2^{\rm T}(q_-)-1>0,\qquad q_\pm\ ^>_<\  E_T/2 \ .  \label {74}
\end{eqnarray}
It is now clear that at the very basic level the breaking of QNS is due to the non-equivelance of the $\phi$ dependences of thermal and shower partons.
Experimental verification of the QNS breaking can therefore render an indirect support for the recombination mechanism involving thermal and shower partons, as we have described.

\section{Conclusion}

At $\sqrt{s_{NN}} = 200$ GeV the density of partons with momentum fraction $x\ ^>_\sim\ 0.03$   is high, and their scattering into $k_T \ ^>_\sim\ 3$ GeV/c can readily occur in a heavy-ion collision.  Thus the formation of ridges due to weak jets is an aspect of the event structure that is pervasive, and should be taken into account in the study of $\phi$ distribution.  Hydrodynamical expansion of the dense medium undoubtedly takes place, but there is no requirement from any fundamental principle that the thermodynamical description must be valid within 1 fm/c after collision.  With semi-hard scattering driving the azimuthal anisotropy, fast equilibration need not take place, but hydrodynamics can still have its realistic application at later time.  Short time-scale physics implies hard or semi-hard processes by uncertainty principle, and it is not the proper domain of equilibrium physics.  We have demonstrated in this study that the observed features of elliptic flow are on the whole reproduced by ridge consideration at low $p_T$ and thermal-shower recombination at intermediate $p_T$.  The validity of hydrodynamics at late time is implicitly incorporated in our approach when the bulk parton distribution is taken to be thermal, i.e., exponential before hadronization.  How to combine semi-hard scattering and hydrodynamical expansion is a time-evolution problem worthy of careful investigation.

Although we have used recombination in all $p_T$ regions, we have shown that quark number scaling is not generally valid.  At low $E_T$ it is approximately valid because the partons that recombine are mostly multiplicative, each having roughly the same $v_2$.  At intermediate $E_T$ the thermal and shower partons contribute to $v_2$ differently, so the mechanism that gives rise to QNS is lost.  

Since rapid thermalization is not required in this study, the partons in the dense medium need not interact more strongly than in the usual formulation of QCD.  Viscosity need not be low, since hydrodynamical calculation should be redone with different initial conditions.  No where in this study suggests  that the medium created in heavy-ion collisions behaves as perfect fluid.

\section*{Acknowledgment}
This work was supported, in part,  by the U.\ S.\ Department of Energy under Grant No. DE-FG03-96ER40972  and by the National Natural Science Foundation in China under Grant No. 10635020 and 10775057 and by the Ministry of Education of China under project IRT0624.

\appendix
\section{Angular Distribution of Ridge Particles}

We describe in this Appendix the subprocesses involved that lead to the distribution of particles in the ridges.  The subject of correlation that depends on the trigger direction is outside the scope of the present problem, and will be described separately in \cite{ch3}.  Our interest here on the single-particle inclusive distribution is less complicated; however, it does require an integration over all angles of the semi-hard partons that generate the ridges.  Thus some aspect of the correlation problem will be adopted here from \cite{ch3}.  Since the problem of angular correlation does not depend critically on the magnitude of $p_T$, we suppress the $p_T$ variable in this Appendix until the very end where contact with the main text is to be made.

It is important to recognize two time scales in this problem:  one is the semi-hard scattering at early time that is sensitive to the initial configuration of the system, and the other is the hadronization process at later time for which the elliptic geometry is more relevant.  For the former the almond-shaped region is bounded by two circular arcs of radius $R_A$ with centers at $x = \pm b/2$, $y = 0$, where the $(x, y)$ coordinates are centered at the origin of the almond with the $x$-axis on the short side and in the reaction plane.  We refer to it as the $A$ geometry.  In this Appendix we normalize all lengths by $R_A$, so the radii are $1$.  For the latter the ellipse is defined by
\begin{eqnarray}
\left({x \over w} \right)^2 + \left({y \over h} \right)^2 = u \ ,
\label{A1}
\end{eqnarray}
where $w = 1 - b/2$, $h = [1 - (b/2)^2]^{1/2}$, and $u = 1$ at initial time.  We refer to this as the $E$ geometry.  We assume that, as the dense system expands, we need only let $u$ increase in using (\ref{A1}) to describe the boundary of the system.  The vector normal to the surface at any point $(x,y)$ is the gradient of $u$, whose azimuthal angle is
\begin{eqnarray}
\psi (x,y) = \tan ^{-1}\left[\left({w \over h} \right)^2 {y \over x} \right]  \ .
\label{A2}
\end{eqnarray}
It specifies the direction of local flow.

Semi-hard scattering can occur at any point inside the almond.  Let it be labeled by $\sf P$ whose coordinates in the $A$ geometry are $(x,y)$.  Limiting ourselves to only the 2D transverse plane on the basis that all particles detected in the final state are in a narrow rapidity bin at $\eta = 0$, we consider one of the created semi-hard partons moving in a direction at azimuthal angle $\phi _s$ toward the boundary, away from the interior.  Let $t$ be the distance between ${\sf P}$ and the boundary measured along the trajectory of the parton.  Because of energy loss to the medium, $t$ cannot be too large if the parton is to emerge and give rise to a ridge in addition to a trigger particle.  In \cite{ch3} a distribution in $t$ is considered with parameters determined by phenomenology.  For our purpose here that is more illustrative than data fitting, we adopt the discrete average value of $t = 0.1$.  That is, we assume that all semi-hard partons are created along the inner circles of radius $r_0 = 0.9$ from either center in the $A$ geometry.  Let us refer to the locus of those points as the {\it rim}.  Thus  ${\sf P}$ is a point on the rim, and can be specified by the angle $\theta$ measured from the pertinent center.  For definiteness, we consider the right half of the almond rim, so $\theta$ varies from $- \Phi_0$ to $+ \Phi_0$, where $\Phi_0=\cos ^{-1}(b/2r_0)$.  Recoil partons that move toward the interior, as well as those that are created there, are totally absorbed and contribute to the bulk.  High momentum jets can get out in any direction, but they are rare and do not contribute to the ridge particles that can influence the azimuthal anisotropy.

The process of energy loss to the medium by the semi-hard parton cannot be calculated reliably.  In \cite{ch3} successive soft emission with probability proportional to the local density is modeled with correlation to the jet trajectory.  Conversion of the lost energy to the enhancement of thermal partons, which subsequently hadronize to form ridge particles, is also hard to treat rigorously.  It is, however, reasonable to assume that for every semi-hard parton at $\phi_s$ there is a corresponding cluster of ridge particles in $\phi$ that has a Gaussian \dis\ around $\phi_s$.  This is essentially a description of what has been observed in ridge phenomenology \cite{jp}, where the trigger direction takes the place of $\phi_s$ here.  More specifically, we consider the correlation function
\begin{eqnarray}
C(x, y, \phi, \phi_s) = D(x, y) G(\phi, \phi_s) \ ,
\label{A3}
\end{eqnarray}
where 
\begin{eqnarray}
G(\phi, \phi_s) = \exp \left[ -(\phi-\phi_s)^2/2\lambda \right]  \ ,
\label{A4}
\end{eqnarray}
\begin{eqnarray}
D(x, y) &=& T_A(s) \left[ 1 - e^{-\sigma T_B (|\vec s-\vec b|)}\right]\\ \nonumber  &+& T_B (|\vec s-\vec b|)\left[1 - e^{-\sigma T_A (s)} \right] \ .\label{A5}
\end{eqnarray}
The last quantity $D(x, y)$ is the local nuclear density at $(x, y)$, related in the Glauber model to the thickness function $T_A (s)$ and $pp$ inelastic cross section $\sigma$ in the conventional way that will not be detailed here.  We are not concerned with the overall normalization because the aim of this Appendix is to derive the $\phi$ dependence of the collected ridge particles generated by semi-hard partons at all points along the rim, and the normalizations of the ridge versus bulk are considered along with the $p_T$ dependence in Sec.\ III.  The parameter $\lambda$ in (\ref{A4}) is found in \cite{ch3} to be $0.11$, which we use in the following.  It corresponds to a half-width of about $20^{\circ}$.

The semi-hard parton is scattered to an angle $\phi_s$ randomly, so the inclusive \dis\ of the ridge particles without trigger must average over all $\phi_s$.  This should be done for each point {\sf P} on the rim.  A mathematically simple and physically reasonable approximation of the result of that averaging is that the average direction of the ridge particle, $\bar{\phi}$, is normal to the surface, since that is the only direction in the problem.  That is the approximation used in \cite{rh2}, where only the $A$ geometry is considered.  Here we treat the problem more quantitatively by averaging over $\phi_s$ at each $(x, y)$ point and then integrating over all points along the rim.  Furthermore, we improve the calculation by using $E$ geometry, since hadrons are formed later when the system develops elliptical shape and they follow the direction of the local flow.  Although $G(\phi, \phi_s)$ in (\ref{A4}) makes no reference to the system shape, the averaging process is sensitive to it.  That is shown explicitly as follows
\begin{eqnarray}
{dN \over d\phi}(x, y) =  {1 \over 2}  \int^{\psi (x,y)+1}_{\psi (x,y)-1}d \phi_s C(x, y, \phi, \phi_s) \  ,
\label{A6}
\end{eqnarray}
where the range of integration is $\pm1$ around the normal to the elliptic surface $\psi (x,y)$, given in (\ref{A2}), which specifies the direction of flow at $(x,y)$ near that surface. We have considered wider range, but the difference from the above is negligible.  The averaging in (\ref{A6}) accounts for the semi-hard partons that move toward the surface and emerge from it.  The difference between the surfaces in the $A$ and $E$ geometries becomes significant, when {\sf P} is near the tips of the almond region.  However, the density is lower there, so the effect of the difference on the yield is not prominent.  

\begin{figure}[tbph]
\includegraphics[width=0.52\textwidth]{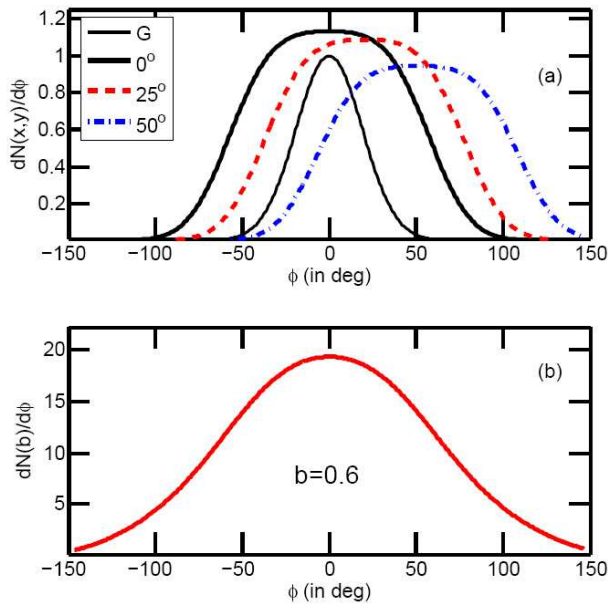}
\caption{(Color online) (a) Azimuthal distributions of ridge particles due to semi-hard partons created at three points on the rim at $\theta=0^{\circ}$ (solid), $\theta=25^{\circ}$ (dashed), and $\theta=50^{\circ}$ (dash-dotted) for $b=0.6 R_A$. Thin solid line is  the Gaussian correlation function $G(\phi, \phi_s)$ for $\phi_s=0^{\circ}$. (b) Azimuthal distributions of ridge particles after integration over all points on the rim. $b$ is labeled in unit of $R_A$.}
\end{figure}

To see the results at various points on the rim, we show in Fig.\ 12(a) 
the azimuthal \dis\ of the ridge particles for $b = 0.6$ created by semi-hard partons, originated at 
three illustrative points, whose azimuthal angles, measured from the center of the right half of the rim, are denoted by $\theta$, where
 $\theta = \tan^{-1}[y/(x+b/2)]$ .  The thin solid line exhibits the shape of  $G(\phi, \phi_s)$ for $\phi_s = 0$.  The thicker solid line shows $dN(x,y)/d\phi$ for $\theta = 0$.  Evidently, the averaging process widens the $\phi$ \dis, since jets at any angle in the range $0 < |\phi_s| < 1$ can contribute.  The other two curves in that figure show the contribution from points at $\theta = 25^{\circ}$ (dashed) and $50^{\circ}$ (dash-dotted).  Their peaks are lower because of the $D(x, y)$ factor in (\ref{A3}) and are centered around $\phi=\theta$.  The experimental detector cannot distinguish the different points of semi-hard scattering, so we must integrate over all $\theta$
\begin{eqnarray}
{dN \over d\phi}(b) =  \int^{\Phi_0 (b)}_{-\Phi _0 b)}d \theta {dN \over d\phi}\left(\theta(x,y)\right)  \ .
\label{A7}
\end{eqnarray}
The result for $b = 0.6$ is shown in Fig.\ 12(b).  We note that the long tail on either side extends well beyond $\phi = \pm 90^{\circ}$.  It is a consequence of the $E$ geometry that we have used, since the normal to the ellipse near the top can be $\psi \approx \pi/2$.  Thus at the tip of the rim hadrons formed from semi-hard partons on the two sides of the rim can overlap, resulting in doubling the value of $dN/d\phi|_{\phi = \pi/2}$ that is calculated from (\ref{A7}), which integrates only the contribution from the right side of the rim.

\begin{figure}[tbph]
\vspace*{0.2cm}
\includegraphics[width=0.52\textwidth]{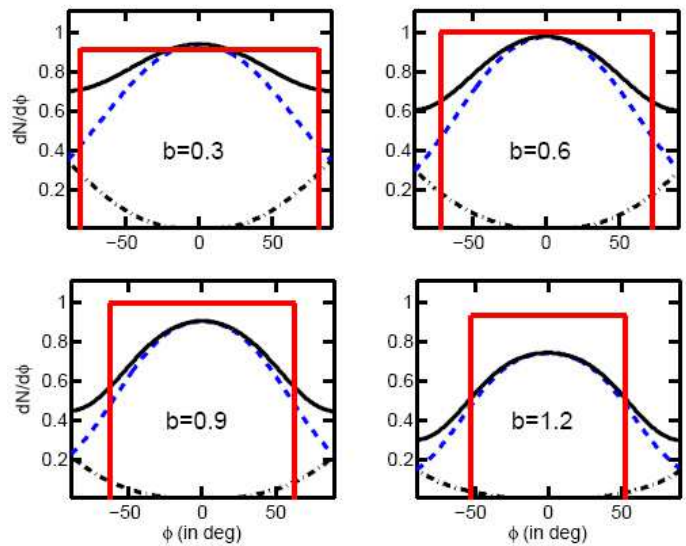}
\caption{(Color online) Azimuthal dependences of ridge particles for $b=0.3, 0.6, 0.9\,  {\rm and} 1.2$ in units of $R_A$. Solid (black) line is the sum of the contributions from the right-side rim (dashed line) and the left-side rim (dash-dotted line). Straight (red) line is the box approximation with width at $|\phi|=\Phi(b)$. Vertical scale is adjusted such that the box height for $b=0.6$ is 1; scales for all other panels are not readjusted.}
\end{figure}

In Fig.\ 13 we show the $\phi$ \dis s for four values of $b$ in the range $-\pi/2 \leq \phi \leq \pi/2$.  In each panel the main contribution (in dashed line) comes from the right side of the rim where $\theta$ is between $-\Phi_0(b)$ and $+\Phi_0(b)$ as seen in (\ref{A7}).  The dashed-dotted lines show the contributions from the left side of the rim --- from $\theta>\pi-\Phi_0(b)$ as well from $\theta<\pi+\Phi_0(b)$ measured from the center of the left half of the rim.  The sum is shown in solid (black) line.  We have checked that at $b = 0$ the \dis\ is absolutely flat, as it should be. The normalization will be discussed presently.

For the purpose of rendering simple mathematics and transparent physics for the determination of $v_2 (p_T)$ in Secs.\ II and III, we now approximate ${dN \over d\phi}(b)$ shown in Fig.\ 13 by step function shown in solid straight (red) line, whose width  is given by Eqs.\ ({1}) and  ({2}).  The values of $\Phi (b)$ for $b = 0.3$, 0.6, 0.9, and 1.2 are $81.4^{\circ}$, $72.5^{\circ}$, $63.3^{\circ}$ and $53.1^{\circ}$, respectively.  
The height of each inverted box is 
determined by matching the area under the box with the area under the calculated ridge curve, which follows from
 (A3) through (A7). We  have given those equations arbitrary overall normalization because  the true normalization of the ridge \dis, $R(p_T, \phi)$, involves the consideration of $p_T$ dependence, which is discussed in Sec.\ III. 
 In Fig.\ 13 we have adjusted the vertical scale so that the height of the box approximation for $b=0.6$ is 1. The relative heights among the different panels of that figure are not adjustable. Thus Fig.\ 13 exhibits clearly the comparison between the results of the calculation of the $\phi$ dependence and the box approximations for various values of $b$, showing that the heights of the boxes remain essentially the same and that the widths specified by $\Phi (b)$ summarize the effective dependence on centrality.
In the lower-right panel of Fig.\ 13 for  $b=1.2$  the two wings of the yield curve extend considerably beyond the box because of the difference between the $A$ and $E$ geometries when $\phi$ is near $\pi/2$. The rapid descends of the wings are acceptably represented by the narrowing of the width of the box.

Although details in each subprocess considered in this Appendix can be made more elaborate with more parameters, the general property of the outcome is clear and cannot deviate too much from what is shown in Fig.\ 13.  The box approximation given in Eq.\ ({4}) captures the essence of the $\phi$ dependence of the ridge yield, $R(p_T, \phi)$, and facilitates the derivation of the very simple, analytic formula, Eq.\ ({8}), for $v_2(p_T, \phi)$.

%\newpage

\end{document}